\documentclass[10pt,twocolumn,prl,aps,showpacs,superscriptaddress,floatfix,
preprintnumbers,reprint,nofootinbib]{revtex4-1}
\usepackage{color}
\usepackage{epsfig}
\usepackage{amsmath,bm}
\renewcommand\({\left(}
\renewcommand\){\right)}

\renewcommand\]{\right]}

\newcommand{\bp}{{\bf p}}
\newcommand{\bv}{{\bf v}}
\newcommand{\br}{{\bf r}}
\newcommand{\bk}{{\bf k}}
\newcommand{\bK}{{\bf K}}

\newcommand{\bF}{{\bf\Phi}}
\newcommand{\bL}{{\bf\Lambda}}
\newcommand{\GF}{G_{\rm F}}

\newcommand{\half}{{\textstyle\frac{1}{2}}}

\newcommand{\st}{{\rm s}_\theta}
\newcommand{\ct}{{\rm c}_\theta}

\newcommand{\sv}{{\rm s}_\varphi}
\newcommand{\cv}{{\rm c}_\varphi}

\begin{document}

\title{Fast Pairwise Conversion of Supernova Neutrinos: A Dispersion-Relation Approach}

\preprint{INT-PUB-16-023, MPP-2016-266}

\author{Ignacio Izaguirre}

\affiliation{Max-Planck-Institut f\"ur Physik
(Werner-Heisenberg-Institut), F\"ohringer Ring 6, 80805 M\"unchen,
Germany}

\author{Georg Raffelt}

\affiliation{Max-Planck-Institut f\"ur Physik
(Werner-Heisenberg-Institut), F\"ohringer Ring 6, 80805 M\"unchen,
Germany}

\author{Irene Tamborra}

\affiliation{Niels Bohr International Academy, Niels Bohr Institute,
Blegdamsvej 17, 2100 Copenhagen, Denmark}

\date{11 January 2016}

%%%%%%%%%%%%%%%%%%%%%%%%%%%%%%%%%%%%%%%%%%%%%%%%%%%%%%%%%%%%%%%%%%%%%%%%%%%%%%%

\begin{abstract}
Collective pair conversion $\nu_e\bar\nu_e\leftrightarrow
\nu_{x}\bar\nu_{x}$ by forward scattering, where $x=\mu$ or $\tau$,
may be generic for supernova neutrino transport. Depending on the
local angular intensity of the electron lepton number carried by
neutrinos, the conversion rate can be ``fast,'' i.e., of the order of
$\sqrt{2}\GF(n_{\nu_e}{-}\,n_{\bar\nu_e})\gg\Delta m^2_{\rm atm}/2E$.
We present a novel approach to understand these phenomena: a
dispersion relation for the frequency and wave number $(\Omega,\bK)$
of disturbances in the mean field of $\nu_e\nu_x$ flavor
coherence. Run-away solutions occur in ``dispersion gaps,'' i.e., in
``forbidden'' intervals of $\Omega$ and/or $\bK$ where propagating
plane waves do not exist.  We stress that the actual solutions also depend
on the initial and/or boundary conditions, which need to be further
investigated.
\end{abstract}

%\pacs{14.60.Pq, 97.60.Bw}

\maketitle

%%%%%%%%%%%%%%%%%%%%%%%%%%%%%%%%%%%%%%%%%%%%%%%%%%%%%%%%%%%%%%%%%%%%%%%%%%%%%%%
% Introduction
%%%%%%%%%%%%%%%%%%%%%%%%%%%%%%%%%%%%%%%%%%%%%%%%%%%%%%%%%%%%%%%%%%%%%%%%%%%%%%%

{\em Introduction.}---The physics of core-collapse supernova (SN) explosions
and neutron-star (NS) mergers raises unique questions about flavor evolution
in environments where neutrinos are dense. Their
decoupling strongly  depends on
flavor because $\beta$ reactions dominate for $\nu_e$ and $\bar\nu_e$. As a
result, the $\nu_e\bar\nu_e$ flux of the SN accretion phase exceeds
the $\nu_x\bar\nu_x$ fluxes \cite{Mirizzi:2015eza}, an effect that is even
more pronounced in NS mergers \cite{Foucart:2015gaa,Malkus:2015mda}.
Moreover, the SN $\nu_e$ flux is larger than the $\bar\nu_e$ one
(deleptonization) and the other way round in NS mergers.

The subsequent flavor evolution matters because SN neutrinos not only carry
away energy, but also deposit some of it in the gain region below the stalled
SN shock by $\nu_e+n\to p+e^-$ and $\bar\nu_e+p\to n+e^+$, thus driving the
delayed explosion. At later stages, neutrinos regulate the nucleosynthesis
outcome in the neutrino-driven wind. The neutrino signal
from the next nearby SN will also depend on the flavor ratio.

In the SN region of interest, the matter density is large and suppresses
conventional flavor conversion of the type $\nu_e(\bp)\to\nu_x(\bp)$, which is
driven by neutrino masses and mixing. This effect becomes important only at
larger radii where neutrinos undergo an MSW resonance~\cite{Mikheev:1986if}.
Stochastic density variations from turbulence might stimulate flavor
conversions \cite{Patton:2014lza}, but have been found to be ineffective
during the accretion phase~\cite{Kneller:2016}.

Neutrino-neutrino interactions can famously change this picture
\cite{Mirizzi:2015eza, Duan:2006an, Duan:2009cd, Duan:2010bg, Chakraborty:2016yeg, Hannestad:2006nj, Raffelt:2007yz, Fogli:2008pt, Pehlivan:2011hp, Volpe:2013jgr}
because flavor off-diagonal refraction by $\nu_e\nu_x$ coherence spawns conversion
\cite{Pantaleone:1992eq, Samuel:1993uw, Duan:2005cp}. In this way,
neutrinos feed back
upon themselves and can develop collective run-away modes.
Neutral-current interactions preserve flavor, so we are dealing with flavor
exchange of the type
\hbox{$\nu_e(\bp)+\nu_x(\bk)\leftrightarrow\nu_x(\bp)+\nu_e(\bk)$} and
especially
\hbox{$\nu_e(\bp)+\bar\nu_e(\bk)\leftrightarrow\nu_x(\bp)+\bar\nu_x(\bk)$}
by forward scattering. Such pairwise swaps preserve net flavor, but
still modify subsequent charged-current interactions.

The impact of refractive $\nu_e\bar\nu_e\leftrightarrow \nu_x\bar\nu_x$
conversion has never been studied in SN simulations because such effects
seemed to arise only beyond the shock wave~\cite{Dasgupta:2011jf}. Yet, Sawyer
has long held that such conclusions result from overly simplified assumptions about
neutrino distributions \cite{Sawyer:2005jk, Sawyer:2008zs, Sawyer:2015dsa}
and recently, other authors have followed suit~\cite{Chakraborty:2016lct,
Dasgupta:2016dbv}. The key issue is the $\nu_e$ and $\bar\nu_e$ angle
distributions to be sufficiently different, in contrast to the traditional
``bulb'' emission model.  Another option is a ``backward'' $\nu_e$ and
$\bar\nu_e$ flux which is unavoidable in the SN decoupling region and also
at larger distances \cite{Cherry:2012zw, Sarikas:2012vb}. The growth rate for
``fast multi-angle instabilities'' is of the order of
\begin{equation}\label{eq:fastrate}
\Phi_0=\sqrt{2}\GF\(n_{\nu_e}{-}\,n_{\bar\nu_e}\)=6.42~{\rm m}^{-1}\,
\frac{n_{\nu_e}{-}\,n_{\bar\nu_e}}{10^{31}~{\rm cm}^{-3}}\,.
\end{equation}
Notice that we use natural units with $\hbar=c=1$ where
$6.42~{\rm m}^{-1}=1.92\times10^{9}~{\rm s}^{-1}=1.27~\mu{\rm eV}$.
This rate is ``fast'' in that it far exceeds the vacuum oscillation
frequency $\Delta m^2_{\rm atm}/2E=0.5~{\rm km}^{-1}$ where we have used $\Delta
m^2_{\rm atm}=2.4\times10^{-3}~{\rm eV}^2$ and $E=12.5~{\rm MeV}$. Fast\break
flavor conversion does not require neutrino masses or mixing, except for
providing seed perturbations. Moreover, energy drops out, forestalling the
characteristic energy-dependent flavor swaps found in many scenarios of
collective flavor conversion~\cite{Mirizzi:2015eza, Duan:2010bg}. More likely,
some sort of flavor equilibration by chaotic evolution of many
nonlinearly coupled modes will occur  \cite{Sawyer:2005jk,
  Sawyer:2008zs, Sawyer:2015dsa,
Hansen:2014paa, Mirizzi:2015fva, Dasgupta:2015iia, Capozzi:2016oyk}.

We here propose a new perspective that vastly simplifies both the conceptual
understanding and the practical treatment of these phenomena. The starting point
is the mean field of $\nu_e\nu_x$ coherence, essentially the off-diagonal
element of the usual $\varrho(t,\br,\bp)$ flavor matrix, which normally
evolves purely kinematically.
However, after including $\nu\nu$ refraction, $\varrho$ becomes
dynamical and we can think of the neutrino medium as supporting flavor waves
described by a wave four vector $K=(\Omega,\bK)$ and a
corresponding polarization vector.

A propagating mode is a collective
disturbance with a certain frequency $\Omega$. To fulfill the equation of
motion (EOM), $\Omega$ may be required to be complex for some $\bK$, leading to
solutions which grow or shrink exponentially in time. Conversely, some
$\Omega$ specified at the boundary may require complex $\bK$ and thus,
exponential solutions as a function of distance. Moreover, various recently
discovered symmetry breaking effects \cite{Sarikas:2012ad, Raffelt:2013rqa,
Raffelt:2013isa, Mangano:2014zda, Duan:2014gfa, Chakraborty:2015tfa} simply
correspond to complex $\bK$ in directions other than the symmetry
axis of the neutrino medium, and/or to different polarizations of our flavor
waves. We here focus on fast modes because they are less familiar, yet may
dominate in environments where previously no conversion was thought to occur.

%%%%%%%%%%%%%%%%%%%%%%%%%%%%%%%%%%%%%%%%%%%%%%%%%%%%%%%%%%%%%%%%%%%%%%%%%%%%%%%
% Mean field of flavor coherence
%%%%%%%%%%%%%%%%%%%%%%%%%%%%%%%%%%%%%%%%%%%%%%%%%%%%%%%%%%%%%%%%%%%%%%%%%%%%%%%

{\em Mean field of flavor coherence.}---We describe the neutrino mean field
by the usual density matrices $\varrho$. For two flavors, we write in the
weak-interaction basis
\begin{equation}
\varrho=\frac{f_{\nu_e}+f_{\nu_x}}{2}+\frac{f_{\nu_e}-f_{\nu_x}}{2}\,
\begin{pmatrix}s&S\\S^*&-s\end{pmatrix}\,,
\end{equation}
where $f_{\nu_e}$ and $f_{\nu_x}$ are the initial occupation numbers. The
complex scalar field $S_\bp(t,\br)$ represents $\nu_e\nu_x$ flavor coherence
for mode $\bp$, whereas the real field $s_\bp(t,\br)$ obeys
$s_\bp^2+|S_\bp|^2=1$ and provides the survival probability by $\half(1+s)$.
 We use the ``flavor isospin convention,'' where $\bar\nu$ has negative energy and
negative $\varrho$, so
the $\bar\nu$ coefficients are $-(f_{\bar\nu_e}{+}f_{\bar\nu_x})/2$ and
$-(f_{\bar\nu_e}{-}f_{\bar\nu_x})/2$.

The usual EOM is
\hbox{$(\partial_t+\bv\cdot{\bm\nabla}_{\br})\varrho=i[\varrho,{\sf H}]$}, where
we ignore collisions \cite{Sigl:1992fn, Cardall:2007zw} and where
the Liouville operator accounts for free streaming. The Hamiltonian matrix is ${\sf H}
={\sf M}^2/2E+v^\mu\Lambda_\mu\,\half\sigma_3+\sqrt{2}\GF\int d\Gamma^\prime\,
v^\mu v^\prime_\mu \varrho^\prime$,
where $\sigma_3$ is a Pauli matrix. The neutrino mass-square matrix ${\sf
M}^2$ is what drives oscillations because it is not diagonal in the weak
interaction basis. The second term is the usual matter effect, where
$v^\mu\Lambda_\mu=\Lambda_0-\bv\cdot{\bL}$, $v^\mu=(1,\bv)$ is the
neutrino four velocity, and $\Lambda_0=\sqrt{2}\GF(n_e{-}\,n_{\bar e})$, with
${\bL}$ the corresponding current. The third term is an integral over
the neutrino phase space, extending to negative energies to include
antineutrinos.

We here study fast modes and thus, dismiss ${\sf M}^2$. As neutrinos are
produced in flavor states, any $\varrho$ matrix beginning and staying
diagonal is a fixed-point solution. Our task is to determine when this fixed
point is stable or unstable. To this end, we use $|S|\ll 1$ and observe that
to linear order $s=\sqrt{1-|S|^2}=1$. Moreover, the EOM no longer depends on
$E$, so we only deal with angle modes described by $\bv$. The same $S_\bv$
applies to $\nu$ and $\bar\nu$, so we only need the angle distribution
of electron lepton number (ELN) carried by neutrinos, which we express as
\begin{equation}
G_\bv=\sqrt{2}\GF\int_0^\infty\frac{dE\,E^2}{2\pi^2}\,
\bigl[f_{\nu_e}(E,\bv)-f_{\bar\nu_e}(E,\bv)\bigr]\,.
\end{equation}
If the $\nu_x$ and $\bar\nu_x$ distributions are not equal, we must
include
$-\[f_{\nu_x}(E,\bv)-f_{\bar\nu_x}(E,\bv)\]$. The ELN potential is
$\Phi_0=\int d\Gamma\,G_\bv$ and the current is ${\bf\Phi}=\int
d\Gamma\,G_\bv\,\bv$. The phase-space integration is over the unit sphere:
$\int d\Gamma=\int d\bv/4\pi$.
We may use coordinates with
$z$ along the radial direction and angles
$(\theta,\varphi)$ to express $\bv=(v_x,v_y,v_z)=(\st\cv,\st\sv,\ct)$,
where $\ct=\cos(\theta)$ and so on.

Assuming that in our test volume, the occupation numbers as well as the matter
density are homogeneous and stationary, the linearized EOM is
\begin{equation}\label{eq:EOM1}
i\(\partial_t+\bv\cdot{\bm\nabla}_{\br}\)S_\bv=
v^\mu(\Lambda+\Phi)_\mu S_\bv- \int d\Gamma'\,v^\mu v'_\mu\,G_{\bv'}S_{\bv'}\,.
\end{equation}
Here $v^\mu(\Lambda+\Phi)_\mu=\Lambda_0+\Phi_0-\bv\cdot(\bL+\bF)$ is the energy shift
due to {\emph{matter and neutrinos}} and $v^\mu v'_\mu=(1-\bv\cdot\bv')$. For a plane wave
$S_\bv(t,\br)=Q_\bv(\Omega,\bK)\,e^{-i(\Omega t-\bK\cdot\br)}$, the~EOM~is
\begin{equation}\label{eq:EOM2}
v^\mu k_\mu\,Q_\bv=-\int d\Gamma'\,v^\mu v'_\mu\,G_{\bv'}Q_{\bv'}\,,
\end{equation}
where $k=K-(\Lambda+\Phi)$ with $k^\mu=(\omega,\bk)$ and $K=(\Omega,\bK)$. Notice that our $\omega$
does not denote $\Delta m^2_{\rm atm}/2E$.

The dispersion relation will be for $(\omega,\bk)$ and depends only on
$G_\bv$. Matter enters through the constant shift $(\Omega,\bK)\to
(\omega,\bk)$ which means going to a rotating frame in flavor space
\cite{Duan:2005cp, Abbar:2015fwa, Dasgupta:2015iia}. $K$ and $k$ have the
same imaginary part, if any. The shift amounts to a global gauge
transformation $S_\bp(r)\to S_\bp(r) e^{i(\Lambda+\Phi)r}$. For the $\varrho$
matrices, it is a global SU(2) gauge transformation.

%%%%%%%%%%%%%%%%%%%%%%%%%%%%%%%%%%%%%%%%%%%%%%%%%%%%%%%%%%%%%%%%%%%%%%%%%%%%%%%
% Dispersion relation
%%%%%%%%%%%%%%%%%%%%%%%%%%%%%%%%%%%%%%%%%%%%%%%%%%%%%%%%%%%%%%%%%%%%%%%%%%%%%%%

{\em Dispersion relation (DR).}---Without $\nu\nu$ interactions,
Eq.~(\ref{eq:EOM2}) implies $v^\mu k_\mu=0$. This purely kinematical relation means
that a spatial disturbance of mode $\bv$ is carried by the Liouville flow,
causing a local time variation with $\omega=\bv\cdot\bk$. Including
$\nu\nu$ interactions, the EOM becomes dynamical. Physically, the local time
variation ``observed'' by another neutrino can lead to a parametric resonance
%\cite{Raffelt:2008hr}
and thus, to run-away solutions.

The right hand side of Eq.~(\ref{eq:EOM2}) has the form $v^\mu a_\mu$
with a ``polarization vector''
$a_\mu=-\int d\Gamma\,v_\mu\,G_{\bv}Q_{\bv}$, so $Q_\bv=v^\mu a_\mu/v^\mu k_\mu$. Insertion
on both sides of Eq.~(\ref{eq:EOM2}) yields $v^\mu a_\mu=-\int
d\Gamma'\,v^\mu v'_\mu G_{\bv'}\, a^\mu v'_\mu/k^\mu v'_\mu$.
Using the metric $\eta^{\mu\nu}={\rm diag}(+,-,-,-)$,
this EOM is $v_\mu\Pi^{\mu\nu}a_\nu=0$. Here the ``polarization tensor''
\begin{equation}\label{eq:Pi}
\Pi^{\mu\nu}=
\eta^{\mu\nu}+\int \frac{d\bv}{4\pi}\,G_\bv\,
\frac{v^\mu v^\nu}{\omega-\bv\cdot\bk}
\end{equation}
contains all physical information, which derives from the ELN angle
distribution $G_\bv$. The EOM $v_\mu\Pi^{\mu\nu}a_\nu=0$ applies to any mode
$v_\mu$ and thus, amounts to
\begin{equation}\label{eq:Pi-a}
\Pi^{\mu\nu}a_\nu=0\,.
\end{equation}
The latter has nontrivial solutions for ${\rm det}\,[\Pi^{\mu\nu}(k)]=0$, providing the
DR. Once we have found solutions $k^{\mu}=(\omega,\bk)$ we can
identify the corresponding polarization vector $a^\mu$ and the eigenfunction
$Q_\bv=a^\mu v_\mu/k^\mu v_\mu$.

To find propagating modes with real $k$, we first pick a direction $\bf \hat
k$ and write $\bk={\bf\hat k} n\omega$ in terms of the refractive index $n$.
In Eq.~(\ref{eq:Pi}), we now pull $1/\omega$ out of the integral and recognize
that ${\rm det}\,[\Pi^{\mu\nu}]=0$ is a quartic equation for $\omega$ as a function of
$n$; i.e., instead of $n(\omega)$ we find four branches $\omega(n)$. Considering
$\bk(n)={\bf\hat k}\, n\,\omega(n)$, we thus find parametric solutions in the
form $[\omega(n),\bk(n)]$. On the other hand, there is no obvious elegant way
to find complex $\omega$ solutions for real $\bk$ or the other way round,
without searching for roots of ${\rm det}\,[\Pi^{\mu\nu}(k)]=0$.

%%%%%%%%%%%%%%%%%%%%%%%%%%%%%%%%%%%%%%%%%%%%%%%%%%%%%%%%%%%%%%%%%%%%%%%%%%%%%%%
% Generic example
%%%%%%%%%%%%%%%%%%%%%%%%%%%%%%%%%%%%%%%%%%%%%%%%%%%%%%%%%%%%%%%%%%%%%%%%%%%%%%%

{\em Generic example.}---We assume axial symmetry of $G_\bv$ and pick
$\bk$ in the radial direction ($z$). In $\Pi^{\mu\nu}$ all terms
linear in $v_{x,y}$ vanish, so Eq.~(\ref{eq:Pi-a}) yields two
equations for $(a_0,a_z)$, providing
$Q_\bv=(a_0-a_z\ct)/(\omega-k_z\ct)$ where we have used
$\bk=(0,0,k_z)$. These are the bimodal and multi-zenith angle (MZA)
polarizations \cite{Raffelt:2013rqa}, which are axially symmetric.
The diagonal $\Pi^{\mu\nu}$ terms from $v_x^2$ and $v_y^2$ yield
degenerate solutions for $a_{x,y}$ with
$Q_\bv=-(a_x\st\cv+a_y\st\sv)/(\omega-k_z\ct)$, the axial symmetry
breaking multi-azimuth angle (MAA) polarizations.

\begin{figure}[b]
\centering
\includegraphics[width=\columnwidth]{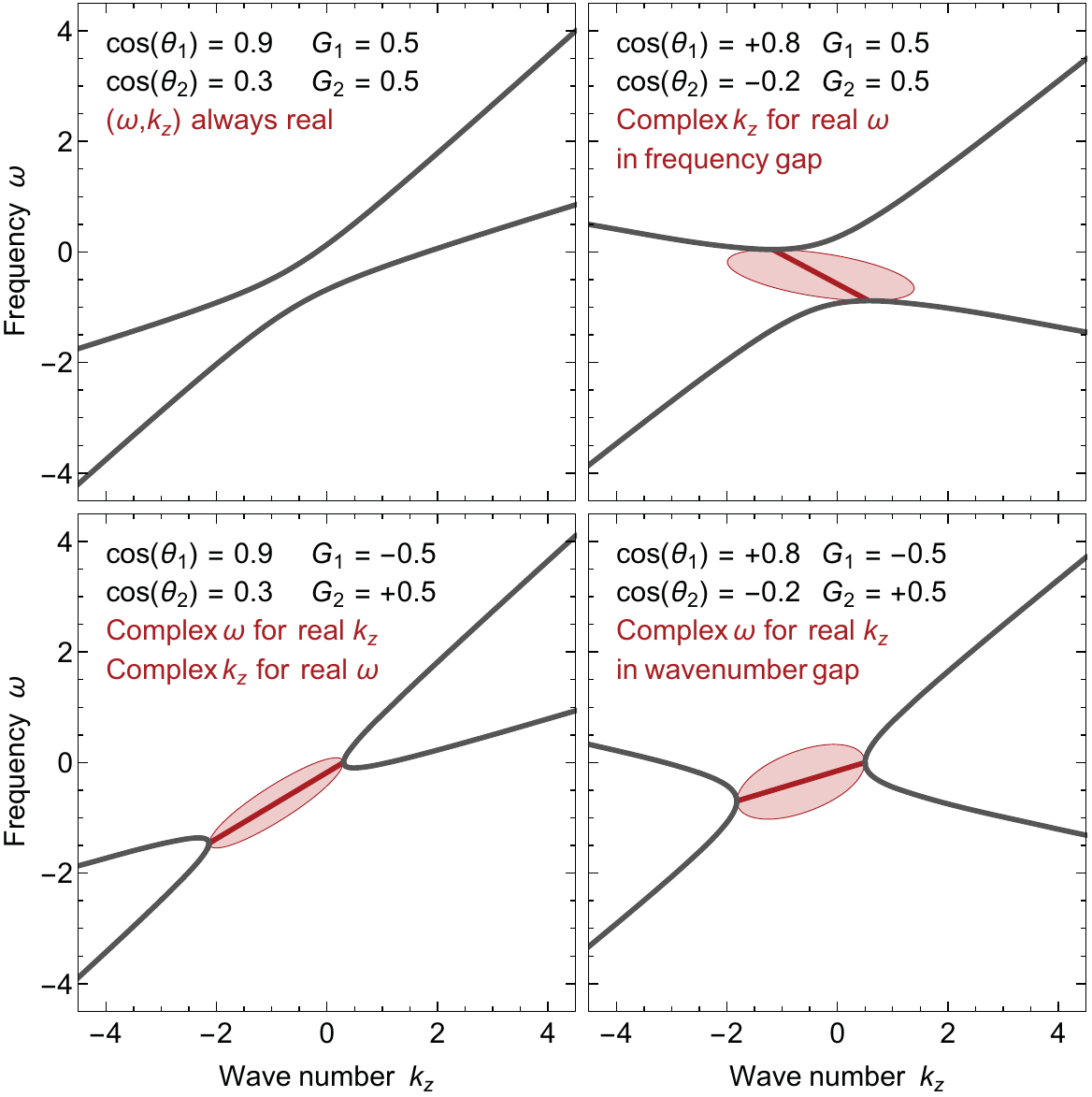}
\caption{Dispersion relations (black lines) for two $\theta$ modes.
The thick red line is ${\rm Re}(\omega)$ for real $k_z$ or ${\rm Re}(k_z)$
for real $\omega$.
The width of the blob is $\pm{\rm Im}(\omega)$ or $\pm{\rm Im}(k_z)$.
{\em Left:} Only outward modes. {\em Right:} One outward and one backward mode.
{\em Top:} Both $\nu_e$ excess. {\em Bottom:} Forward mode
$\bar\nu_e$ excess.}
\label{fig:gaps}
\end{figure}

To be explicit, we study the simplest non-trivial case: two $\theta$ modes
representing two zenith ranges, i.e.,
$G_\bv=G_1\delta(\ct-{\rm c}_1)+G_2\delta(\ct-{\rm c}_2)$.
The axially symmetric polarizations produce a
quadratic form in both $\omega$ and $k_z$, implying that the
DRs are hyperbolas in the $\omega$--$k_z$--plane,
as shown in Fig.~\ref{fig:gaps}. The axially breaking polarizations
provide similar results.

The left panels use forward modes ($0<\cos\theta_{1,2}<1$) as
in traditional bulb
emission. If $\nu_e$ dominate in both modes (upper left), both $\omega$ and
$k_z$ are real: No fast flavor conversion occurs. If one mode has a
$\bar\nu_e$ excess ($G_1<0$), the DR has a gap, providing
complex $\omega$ for real $k_z$ and the other way round as indicated by
the red blob. Disturbances with $k_z$
in the gap grow exponentially in time. A real $\omega$ imposed at the
boundary causes exponential spatial growth. These conclusions carry over to
more general $G(\theta)$ where one needs a crossing from positive to negative
ELN intensities to obtain a dispersion gap, which, in turn,
enables fast flavor conversion, similar to spectral
crossings for slow modes \cite{Dasgupta:2009mg, Fogli:2009rd, Mirizzi:2012wp}.

One forward and one backward mode with $\nu_e$ excess (upper right)
produce two branches of real $\omega$ for all $k_z$, but an $\omega$
gap. All spatial disturbances
propagate, but a ``forbidden'' frequency imposed at the boundary causes
exponential spatial growth. If instead, one of our two modes has $\bar\nu_e$
excess (lower right), there is a gap in $k_z$. Wave numbers in this range imply
temporal run away.

The direction of a general $\bk$ can be chosen such that it feels forward and backward
modes, even if all modes are forward in the SN frame. If $G_\bv>0$
everywhere (no crossing), such cases produce a DR analogous to
the upper right panel (an $\omega$ gap).  The neutrino flow is a
very anisotropic medium, so dispersion strongly depends on ${\bf\hat
  k}$. Moreover, some components of $\bk$ may be real and only one of
them complex, producing exponential variation in only one spatial
direction for a certain $\omega$ gap.

%%%%%%%%%%%%%%%%%%%%%%%%%%%%%%%%%%%%%%%%%%%%%%%%%%%%%%%%%%%%%%%%%%%%%%%%%%%%%%%
% Realistic distribution
%%%%%%%%%%%%%%%%%%%%%%%%%%%%%%%%%%%%%%%%%%%%%%%%%%%%%%%%%%%%%%%%%%%%%%%%%%%%%%%

\begin{figure}[h]
\centering
\includegraphics[width=0.70\columnwidth]{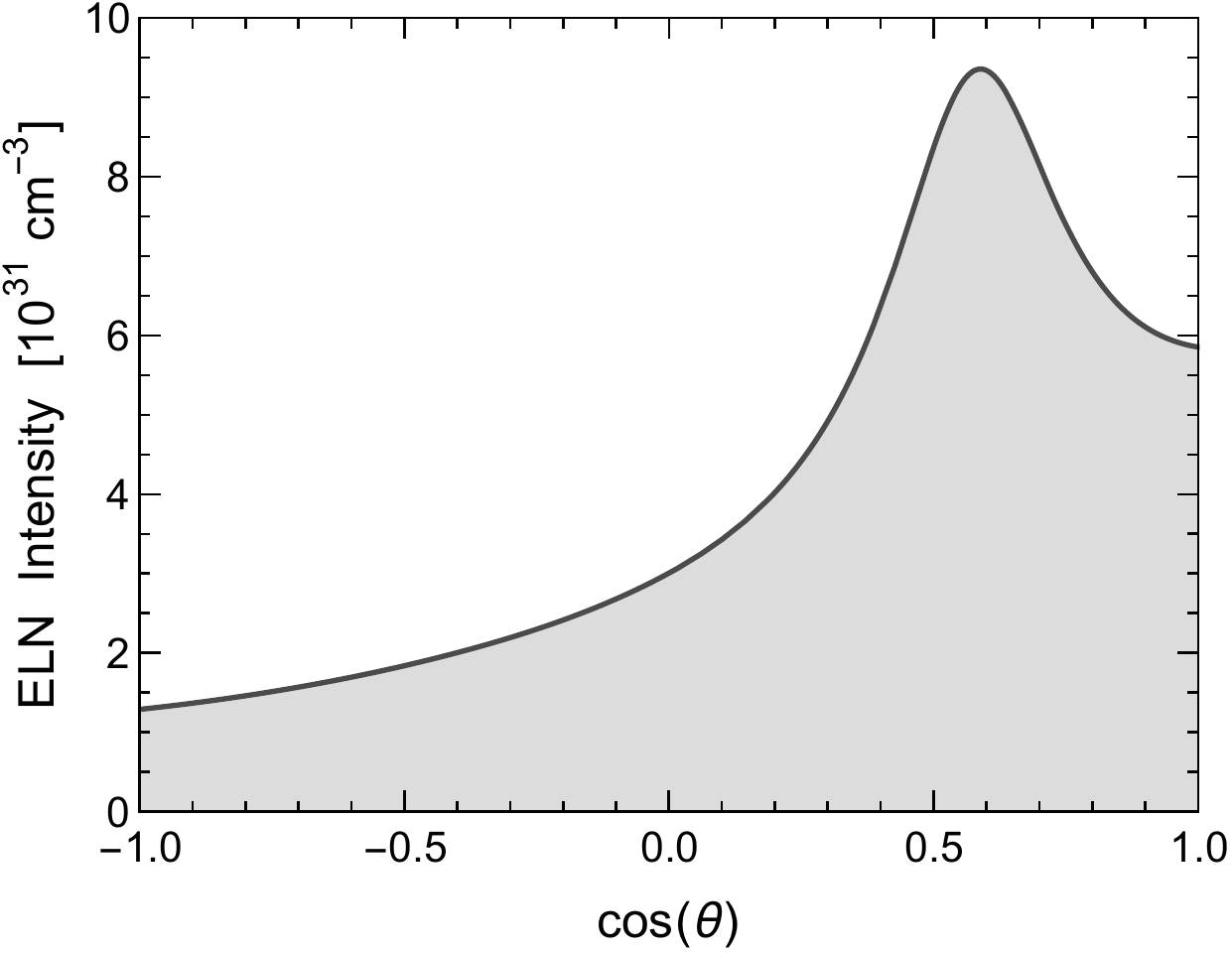}
\caption{Electron lepton number (ELN) angle distribution $G_\theta$ of
  a $15\ M_\odot$ SN simulation at $280~{\rm ms}$ post bounce and a
  radius $37~{\rm km}$. We plot an ELN number density, to be converted
  to a weak potential by Eq.~(\ref{eq:fastrate}). We show a mildly
  smoothed approximation suitable for analytic post
  processing.} \label{fig:ELN-distribution}
\end{figure}

{\em Realistic distribution.}---The flavor-dependent neutrino angle
distributions from SN simulations are not readily
available. To gain intuition, we have extracted the ELN distributions
from a Garching simulation of a $15\ M_\odot$ progenitor~\cite{Sarikas:2012vb,Sarikas:2011am,garc}.
Figure~\ref{fig:ELN-distribution} shows a typical case not far from
the decoupling region. For larger distances, the ELN profile is
horizontally compressed near the forward ($\cos\theta=1$) direction,
although backward modes ($\cos\theta<0$) are never empty. One
key feature is the forward dip due to $\bar\nu_e$ being more
forward peaked than $\nu_e$. However, we have not found any place or
time in this model where this dip would go negative.

\begin{figure}
\centering
\includegraphics[width=0.8\columnwidth]{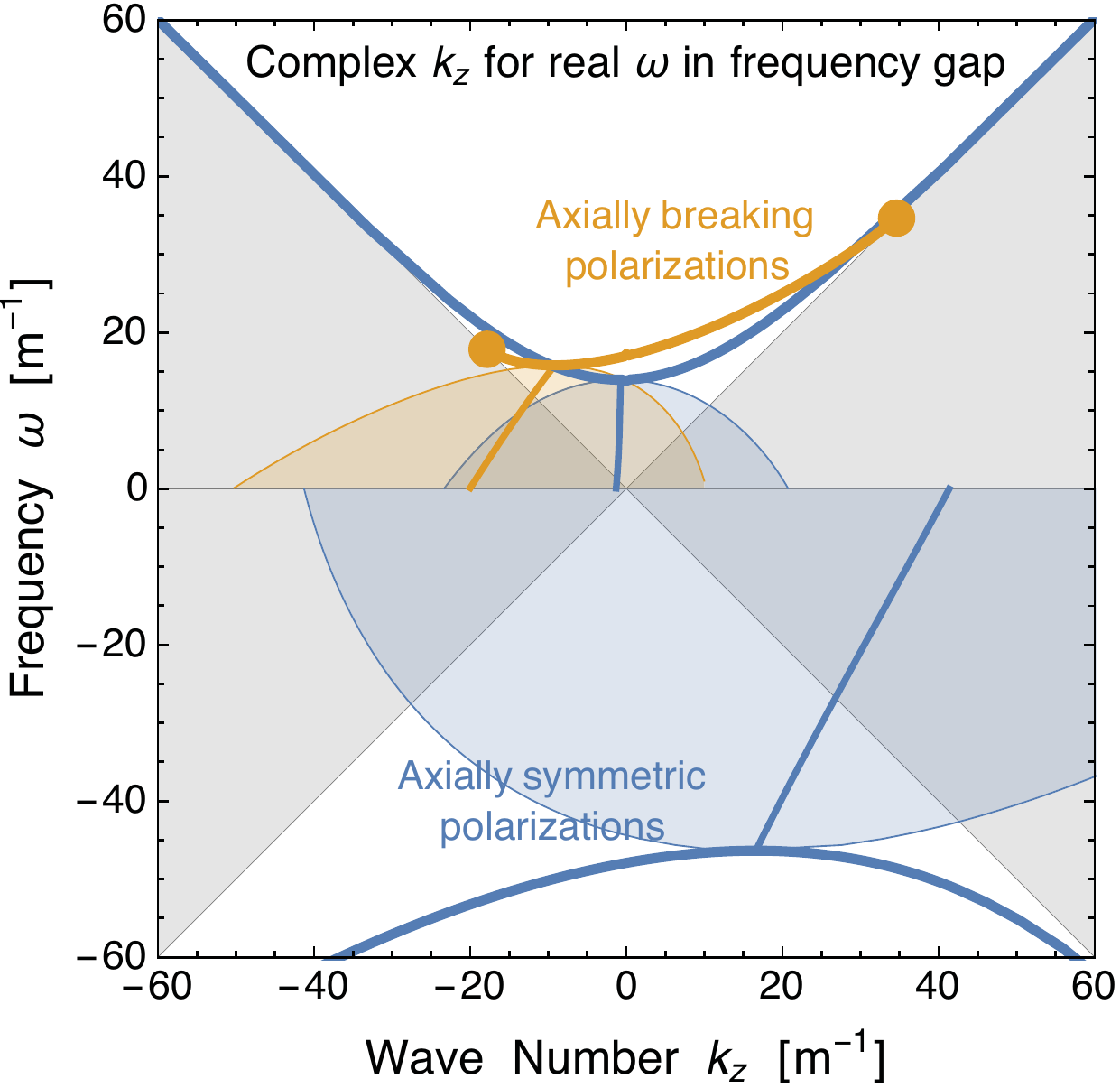}
\caption{Dispersion relations for $\bk=(0,0,k_z)$ with $G_\theta$
  shown in Fig.~\ref{fig:ELN-distribution}. {\em Blue:} Axially
  symmetric polarizations.  {\em Orange}: Two degenerate axially
  breaking polarizations.  {\em Dots:} End of a branch. {\em Filled
    regions:} Complex solutions in analogy to the red blobs in
  Fig.~\ref{fig:gaps}, where the semi-thick solid line is ${\rm Re}(k_z)$
  and the edge of the blob indicates $\pm{\rm Im}(k_z)$.  {\em Gray
    region:} Zone of avoidance for real $(\omega,k_z)$.}
\label{fig:Garching-disp}
\end{figure}

Figure~\ref{fig:Garching-disp} shows the DR implied by
$G_\theta$ of Fig.~\ref{fig:ELN-distribution} for a radial-moving mode with
$\bk=(0,0,k_z)$.  Without $\nu\nu$ interactions, the DR is $\omega=\ct k_z$ for
any angle mode $\ct$ (gray-shaded region). Including
$\nu\nu$ interactions, this region becomes a ``zone of avoidance'' for
propagating collective oscillations as
$Q_\bv\propto 1/(\omega-\ct k_z)$ would be singular.
The thick blue lines are the dispersion
relations for the axially symmetric polarizations.
The two degenerate axially breaking ones (thick orange) end at the big dots on the border
of the zone of avoidance. In the frequency gap, $k_z$ is complex.
We show its real part by semi-thick blue and orange lines
ending on the horizontal axis. In analogy to the red blobs in
Fig.~\ref{fig:gaps}, the blue and orange shaded regions which
kiss the dispersion curves indicate
the imaginary part of $k_z$, i.e., $k_z$ in the frequency gap
has a real part
(semithick line) plus/minus an imaginary part (edge of the blob).

%%%%%%%%%%%%%%%%%%%%%%%%%%%%%%%%%%%%%%%%%%%%%%%%%%%%%%%%%%%%%%%%%%%%%%%%%%%%%%%
% Gap
%%%%%%%%%%%%%%%%%%%%%%%%%%%%%%%%%%%%%%%%%%%%%%%%%%%%%%%%%%%%%%%%%%%%%%%%%%%%%%%

{\em Growth in the gap.}---Any type of ELN distribution probably occurs
somewhere in NS mergers or 3D SN models, but in our 1D model,
$G_\bv$ is always positive and has no crossings. Hence, dispersion is similar to an EM
wave in plasma: For every $\bk$ there is a real $\omega$,
but there is an $\omega$ gap where the EOM requires $\bk$ to be complex.

In analogy to the stability analyses for slow modes~\cite{Banerjee:2011fj},
exponential spatial growth obtains if at an interface (e.g.\ the neutrino sphere),
a forbidden frequency is prescribed. The latter was chosen to be stationary
in the frame where ${\sf M}^2$ is static, i.e.,
$\Omega=\omega+\Lambda_0+\Phi_0=0$, and the system was stable
(real $\bK$) in this region. However, the matter density and neutrino angle distribution evolve
with radius, so a propagating wave can enter a forbidden frequency band.
Exponentially damping and growing
solutions ensue, the latter ones quickly taking over. Beginning with
Ref.~\cite{Duan:2006an}, such exponential growth starting at some
``onset radius'' has been found in many numerical studies.

Notice the difference to EM waves entering a forbidden region, e.g.,
radio waves in the ionized upper atmosphere.  The plasma frequency
prevents propagation and they are reflected---they do not grow
exponentially in the ionosphere.  Flavor waves obey a first-order differential equation,
probably explaining this difference.

Recently it was argued that one should not pick $\Omega=0$ a priori because
every frequency would have {\em some\/} amplitude at the boundary
\cite{Abbar:2015fwa, Dasgupta:2015iia}. In this case
the system is spatially unstable
everywhere if it has a frequency gap.

{\em Boundary conditions.}---However, it is not obvious that the
picture of the flavor field being driven by an external frequency
at some ``neutrino sphere'' is an appropriate description altogether.
Ignoring collisions and without a {\em physical\/} interface, the EOM applies
on both sides of an assumed boundary surface. The
length scales for fast flavor conversion are small, so
something like the traditional
bulb model is not justified in any obvious sense.
Furthermore, the inclusion of backward modes may require to specify
boundary conditions in different
spatial regions. In a SN, all inward moving neutrinos come from neutral-current
scattering of the outward moving ones; hence, inward
and outward flows are flavor-correlated beyond what is prescribed by the EOM.

The DR alone only indicates which solutions are consistent with the EOM,
but not which ones will actually occur. We would be
sure that the system was always stable if the DR did not have any gaps, which,
however, seem to be generic.
Except for quantum fluctuations or hypothetical flavor-violating
interactions~\cite{Dasgupta:2010ae, EstebanPretel:2009is, deGouvea:2012hg},
${\sf M}^2$ is the only source of seed perturbations. However,
which spectrum of flavor disturbances is produced, and where,
remains to be better understood.

%%%%%%%%%%%%%%%%%%%%%%%%%%%%%%%%%%%%%%%%%%%%%%%%%%%%%%%%%%%%%%%%%%%%%%%%%%%%%%%
% Summary
%%%%%%%%%%%%%%%%%%%%%%%%%%%%%%%%%%%%%%%%%%%%%%%%%%%%%%%%%%%%%%%%%%%%%%%%%%%%%%%

{\em Summary.}---We have derived a general dispersion relation (DR)
for disturbances in the mean field of $\nu_e\nu_x$ coherence. This approach
corroborates that
fast run-away solutions can indeed occur as first shown by Sawyer.
We have found that it is the local $\nu_e$ minus $\bar\nu_e$,
angle distribution $G_\bv$, that drives this effect.
Therefore, $G_\bv$ should be investigated
in a larger class of SN models, notably in 3D simulations exhibiting the LESA effect
\cite{Tamborra:2014aua}. The presence of ``crossings'' in $G_\bv$
would signify $\bk$ gaps in the DR and concomitant {\em temporal\/} instabilities,
which depend on the {\em initial\/} conditions of the flavor disturbances.

At present it looks like
$\omega$ gaps are the most generic dispersion form, so the spatial boundary
conditions and their time variation are needed to
understand the generic behavior of the flavor field.
Eventually, one may
not get around, including the collision term in the EOM, to see which modes
of the flavor field are actually excited.

While the DR alone does not prove that fast pairwise flavor
conversion indeed occurs, it may well be a generic phenomenon for SN neutrinos.
The impact of flavor equilibration in the decoupling
region should be phenomenologically explored.
The relevant length scales are much smaller
than the resolution of SN simulations, so one anyway needs
a schematic implementation.
Although the details remain speculative,
non-trivial modifications of shock reheating may be expected.

%%%%%%%%%%%%%%%%%%%%%%%%%%%%%%%%%%%%%%%%%%%%%%%%%%%%%%%%%%%%%%%%%%%%%%%%%%%%%%%
%Acknowledgements %%%%%%%%%%%%%%%%%%%%%%%%%%%%%%%%%%%%%%%%%%%%%%%%%%%%%%%%%%%%%
%%%%%%%%%%%%%%%%%%%%%%%%%%%%%%%%%%%%%%%%%%%%%%%%%%%%%%%%%%%%%%%%%%%%%%%%%%%%%%%

{\em Acknowledgments.}---GR acknowledges partial support by the
Deutsche Forschungsgemeinschaft (grant EXC 153) and the
Horizon-2020 Marie Sk\l{}odowska-Curie Actions of the European Union  (Grant No.\
H2020-MSCA-ITN-2015/674896-ELUSIVES). IT acknowledges support from the
Knud H{\o}jgaard Foundation, the Villum Foundation (Project
No.\ 13164) and the Danish National Research Foundation (DNRF91). We
thank the Institute for Nuclear Theory at the University of Washington for
hospitality and the DOE for partial support during early stages of
this work.

%%%%%%%%%%%%%%%%%%%%%%%%%%%%%%%%%%%%%%%%%%%%%%%%%%%%%%%%%%%%%%%%%%%%%%%%%%%%%%%
%%%  Bibliography  %%%%%%%%%%%%%%%%%%%%%%%%%%%%%%%%%%%%%%%%%%%%%%%%%%%%%%%%%%%%
%%%%%%%%%%%%%%%%%%%%%%%%%%%%%%%%%%%%%%%%%%%%%%%%%%%%%%%%%%%%%%%%%%%%%%%%%%%%%%%

%%%%%%%%%%%%%%%%%%%%%%%%%%%%%%%%%%%%%%%%%%%%%%%%%%%%%%%%%%%%%%%%%%%%%%%%%%%%%%%
\end{document}